\documentstyle[12pt]{article}
\author{U. Kasper, S. Kluske, M. Rainer, \\
\underline{S. Reuter}, H.-J. Schmidt
\vspace{1cm} \\
\small
Universit\"at Potsdam, Mathematisch-naturwiss. Fakult\"at \\
\small
 WIP-Projektgruppe Kosmologie,  D - 14415 POTSDAM, Germany
\normalsize}
\title{Stability properties of the Starobinsky cosmological
model}
\date{}
\begin{document}
\maketitle
\begin{abstract}
We discuss the instabilities appearing in the
cosmological model with a quasi de Sitter phase following from a
fourth-order gravity theory. Both the classical equation as
well as the quantization in form of a Wheeler - De Witt
equation are conformally related to the analogous model with
Einstein's theory of gravity with a minimally coupled scalar
field. Results are: 1. In the non-tachyonic case, classical
fourth-order gravity is not more unstable than Einstein's
theory itself. 2. The well-known classically valid conformal
relation is also (at least for some typical cases) valid on
the level of the corresponding Wheeler - De Witt equations,
which turns out to be a non-trivial statement.
\newline
\newline
{to appear in:
\newline
{\bf Proc. Sem. Relativistic Astrophysics Potsdam 1994},
\newline
Ed .: J. M\"ucket}
\end{abstract}
\section{ Introduction}
The inflationary cosmological model proposed in [1] following
from fourth-order gravity got the name "Starobinsky model" in
1985, cf. [2]. Its stability properties have been discussed
e.g. in [3], [4], and [5]. The results do not fully coincide,
cf. the successive papers [6] and [7]. We shall not repeat
those points that were caused by simple mathematical errors.
However, it seems valuable to sum up that part of the
discussion which originated from different notions of stability, see
sct. 2. \\
The generalization to gravitational field equations of
order higher than fourth, see [8-12],  is discussed in sct. 3.
When quantized, a Wheeler - De Witt equation for the
Starobinsky model appears. From the first glance, one could
conjecture, that the conformal relation between Einstein's theory
with a minimally coupled self-interacting scalar field and
fourth-order gravity can simply be carried over, but Duff [13]
showed that conformal equivalence of classical theories need
not survive in the quantum theory. (For clarity, we mention that Duff
considers a conformally invariant theory, i.e., a theory which is
non-trivially conformally equivalent to itself whereas we consider
two different but conformally equivalent theories; but the argument remains
the same.) So one must look into the
details of the theory, see e.g. Refs. [11, 14-18] and sct. 4.

\section{Different notions of stability}
We consider a Lagrangian $L = L(R)$ with curvature scalar $R$
in a region where
\begin{eqnarray}
\frac{dL}{dR} \frac{d^2 L}{dR^2} \neq 0 \qquad \mbox{.}
\end{eqnarray}
Then one can perform a conformal rescaling of this fourth-order
gravity theory to Einstein`s theory with a minimally coupled
scalar field. In this picture, instabilities in form of ghosts
(wrong sign of the kinetic term) and tachyons (wrong sign of
the mass term) could appear. It holds: ghost never appear,
tachyons sometimes do, but for the Starobinsky model tachyons are
also excluded:
\begin{eqnarray}
L = \frac{R}{2} - \frac{l^2}{12} R^2, \qquad R < \frac{3}{l^2}
\qquad \mbox{.}
\end{eqnarray}
($l \approx 10^{-28} cm $, so the inequality in $(2)$ is not a
real restriction, but one
should mention that in the limit $R \rightarrow
\frac{3}{l^2}$ a new type of instability occurs.)
It turns out, that singularities appearing in
the Starobinsky model are restricted to big bang-type
singularities already known from General Relativity. So one can
conclude [7]: Fourth-order gravity defined by $(2)$ is not more
unstable than General Relativity. \par
A second point of view is the following: General Relativity is
not a renormalizable theory, but it can get this property if
one adds curvature squared terms to the action. However it does
not suffice to add $R^2$, one also needs the square of the
Weyl tensor. The result is: the absence of both ghosts and
tachyons requires the corresponding coefficient to be
artificially fine-tuned to zero. \par
A third point of view is outlined in [5]: One could interpret
the $R^2$-term in $(2)$ to mimic some quantum gravitational
effects. But then the interpretation of the solution changes:
one considers the spatially flat Friedman model, solves the
classical vacuum equation following from $(2)$ and gets
expanding solutions having the property that the Hubble parameter
remains in a strip arbitrarily close to zero for an
arbitrarily long amount of time. So, by the usual quantum fluctuations
one should expect that there is a large probability for the universe
to have a moment in time, where the Hubble parameter is negative.
This would cause a positive probability for having a recollapse, which
is classically excluded for the spatially flat Friedman model.
Up to this point we agree with Suen [5]. Here we want to
clarify that this argument has nothing to do with the
fourth-order terms: it analogously applies to the Einstein theory and
the corresponding Einstein-de Sitter model of the universe.

\section{Sixth and higher order equations}
{}From the Lagrangian $R \Box R$ a sixth-order field equation
appears. It was considered in [8]. More general, a term with
$R \Box^k R$ in the Lagrangian yields field equations of order
$2k+4$. It turns out [9-12] that the conformal transformation
to Einstein's theory with scalar fields is possible, but it
always yields ghosts. In this sense, higher order theories
give additional instabilities. \\
Nevertheless, it makes sense to ask whether the de Sitter
space-time is an attractor solution in gravity theories of the
above mentioned type. For $k=2$ see [10], for general $k$ the
result is as follows: let $k\geq 1$ and
\begin{eqnarray}
L = \sum_{i=0}^{k} a_i R \Box^i R, \qquad a_k \neq 0
\end{eqnarray}
In the set of spatially flat Friedman models the de Sitter
space-time represents an attractor solution if the
coefficients $a_i$ fulfil suitable (complicated) inequalities, but here it
suffices to mention:
\begin{enumerate}
\item For every value $k$, these inequalities can be fulfilled,  \\
\item If these inequalities are fulfilled, then $a_0 \neq 0$,
i.e., if the $R^2$-term is absent, then inflation is no longer
typical.
\item If the Einstein-Hilbert term is added as $+ \epsilon R$ with
the 'correct' sign (non-tachyonic case), then the de Sitter space-time
becomes a transient attractor with a typical mean life time
(german: 'Verweildauer') of the transient phase
$<\tau> \ \sim \ \vert \epsilon \vert \sp{- \gamma}$ where
$\gamma = \frac {1}{2}$ .
\end{enumerate}
\section{The Wheeler - De Witt equation}
Deriving the Wheeler - De Witt equation for higher order theories
different problems occur. One of the main difficulties is the question
how to express higher order theories in the Hamiltonian formalism, which is
of first order. Here we want to discuss two approaches, one with
and the other without additional constraints. In the first one the
first derivative of the cosmic scale factor
with respect to the time is introduced as a new coordinate [14]. This
definition has to be preserved as a constraint to the system.
One considers gauge transformations and first class constraints.
It is possible that there exist more first class constraints than gauge
symmetries in the Lagrange-function. We expect that total and extended
Hamiltonians (in the sense of the Dirac formalism) become equivalent,
supposed, the Lagrangian is a geometrical one, [16]. \par
The second approach [11, 17, 18] introduces the second derivative
of the cosmic scale factor with respect to
the time as a new coordinate. With this definition it is possible,
at least for the Friedman models (except $k=-1$, open universe) and
Lagrangian $L= f(R)$, where $f$ is an arbitrary function of $R$,
to avoid an additional constraint. It is
intrinsically given by one of the canonical equations. The other
canonical equation is equivalent to the trace of the field equation.
Classically, higher order theories are conformally related to Einstein's
theory with a minimally coupled scalar field. Now one can ask whether
those theories are still conformally related to each other if they were
quantized. This is not a trivial question as Duff [13] has shown
that conformal equivalence does not always survive in quantum theory.
It does, at least under the following conditions for the Friedman model:
\begin{enumerate}
\item $L=f(R)$, and $f(R)$ arbitrarily chosen function of $R$,
\item intrinsical definition for the introduced canonical coordinate,
\item the factor-ordering-problem is solved in that way, that the discussed
Wheeler - De Witt equation is covariant in superspace.
\end{enumerate}
If one is going to study more systematically the dynamics
not only of the isotropic Friedman models, but also of arbitrary homogeneous
models, the non-Hausdorff topology in the set of corresponding Lie algebras
(which are related to the isometry groups of the corresponding cosmological
model) might play a crucial role [19].

\subsection*{}
Here, we presented the results of the Cosmology group of
Potsdam University concerning the Starobinsky model; to get a more
balanced reference list one should also look at the papers
cited in Refs. [1 - 20], especially, let us mention the review
article [20], where the status of the Starobinsky model up to
1992 has been outlined, and an extensive reference list was
given.
\subsection*{Acknowledgements}
\small
The authors gratefully acknowledge financial support from
Deutsche Forschungsgemeinschaft and
 from the Wissenschaftler-Integrations-Programm.
\normalsize
\section{References}

[1] Starobinsky, A. A., Phys. Lett. B {\bf 91} (1980) 99. \newline
[2] Vilenkin, A., Phys. Rev. D {\bf 32} (1985) 2511. \newline
[3] M\"uller, V., Schmidt, H.-J., Gen. Relat. Grav. {\bf 17} (1985)
769. \newline
[4] Starobinsky, A. A., Schmidt, H.-J., Class. Quant. Grav.
{\bf 4}    (1987) 695. \newline
[5] Suen, W., Phys. Rev. D {\bf 40} (1989) 315. \newline
[6] M\"uller, V., Int. J. Mod. Phys. D {\bf 3} (1994) 241. \newline
[7] Schmidt, H.-J., Comment to Ref.[5], Phys. Rev. D in print. \newline
[8] Gottl\"ober, S., Schmidt, H.-J., Starobinsky, A.A., Class.
Quant. Grav. {\bf 7} (1990) 893. \newline
[9] Schmidt, H.-J., Class. Quant. Grav. {\bf 7} (1990) 1023. \newline
[10] Battaglia Mayer, A., Schmidt, H.-J., Class. Quant. Grav.
     {\bf 10} (1993) 2441. \newline
[11] Schmidt, H.-J., Phys. Rev. D {\bf 49} (1994) 6354. \newline
[12] Kluske, S., Schmidt, H.-J., Abstract 13. \"osterr.
    Math.-Kongre\ss\ Linz 1993.  \newline
[13] Duff, M., Class. Quant. Grav. {\bf 11} (1994) 1387. \newline
[14] Kasper, U. Class. Quant. Grav. {\bf 10} (1993) 869. \newline
[15] Kasper, U., Schmidt, H.-J., Abstracts Conf. GR 13,
Cordoba 1992, Ed.: P. Lamberti, p.  564. \newline
[16] Kasper, U., Preprint Uni Potsdam 94/10, subm. to Class. Quant. Grav.
\newline
[17] Reuter, S., Schmidt, H.-J., Proc. Conf. Diff. Geom. and
     Applic. Opava 1992, Ed.: D. Krupka, p. 243. \newline
[18] Kasper, U., Preprint Uni Potsdam 94/11, subm. to Class. Quant. Grav.
\newline
[19] Rainer, M., Thesis Potsdam University 1994, unpublished. \newline
[20] Gottl\"ober, S., M\"uller, V., Schmidt, H.-J., Starobinsky,
     A. A., Int. J. Mod. Phys. D {\bf 1} (1992) 257 - 279.

\end{document}